\documentclass[12pt]{article}
\usepackage{natbib,bm}
\usepackage{amsmath}
\usepackage{amssymb}
\usepackage{graphicx}
\usepackage[dvips]{epsfig}
\usepackage{lscape}

\renewcommand{\baselinestretch}{1.1}
\tiny\normalsize

\def\c#1{\ensuremath{\mathcal{#1}}}

\newcommand{\nmathbf}{\bm}

\def\bfw{\nmathbf w}
\def\bfx{\nmathbf x}

\newcommand{\cfF}{\mbox{\c{F}}}

\newcommand{\cfJ}{\mbox{\c{J}}}

\newcommand{\cfP}{\mbox{\c{P}}}

\def\boldfacefake#1{\kern-4pt
   \hbox{ \mathsurround=0pt
   \hbox to 0.4pt{$#1$\hss}\hbox to 0.4pt{$#1$\hss}\hbox {$#1$}}}

\newcommand{\E}{\mbox{E}}

\newcommand{\Var}{\mbox{Var}}

\newcommand{\ok}{\hfill\fbox{}}

\newcommand{\btable}{\begin{table}[h]\centering}
\newcommand{\etable}{\end{table}}
\newcommand{\bt}{\begin{parag}\small \let\b=\nsb \let\sb=\nssb \begin{tabular}}
\newcommand{\et}{\end{tabular}\let\b=\nb \let\sb=\nsb\end{parag}}

\newenvironment{parag}{\par}{\par}
\newenvironment{dif}
    {\begin{parag}\small \let\b=\nsb \let\sb=\nssb \begin{parag}}
    {\let\b=\nb \let\sb=\nsb \end{parag}\end{parag}}

\newenvironment{proof}{\begin{dif} \noindent{\em Proof.~}}
            {\ok\vspace*{10pt}\end{dif}}
\newenvironment{exa}{\begin{list}{}
           {\setlength{\leftmargin}{10pt}
            \setlength{\rightmargin}{\leftmargin}}
           \item\begin{ex}\em}{\end{ex}\end{list}}

\newcommand{\be}{\begin{eqnarray}}
\newcommand{\ee}{\end{eqnarray}}
\newcommand{\ba}{\begin{eqnarray*}}
\newcommand{\ea}{\end{eqnarray*}}

\newtheorem{theorem0}{Theorem}
\newtheorem{lemma0}{Lemma}
\newtheorem{remark0}{Remark}
\newtheorem{fact0}{Fact}
\newtheorem{example0}{Example}
\newtheorem{definition0}{Definition}
\newtheorem{corollary0}{Corollary}
\newtheorem{proposition0}{Proposition}
\newtheorem{algorithmY}{Algorithm}

\newenvironment{theorem}{\begin{theorem0} \mbox{} }{\end{theorem0}}

\newenvironment{corollary}{\begin{corollary0} \mbox{} }{\end{corollary0}}

\newcommand{\reals}{\mbox{\rm I\kern-.20em R}}
\newcommand{\sreals}{\mbox{\small \rm I\kern-.20em R}}

\newcommand{\bdfn}{\begin{dfn}}
\newcommand{\edfn}{\end{dfn}}
\newcommand{\bteo}{\begin{teo}}
\newcommand{\eteo}{\end{teo}}
\newcommand{\bexa}{\begin{exa}}
\newcommand{\eexa}{\end{exa}}
\newcommand{\bdif}{\begin{dif}}
\newcommand{\edif}{\end{dif}}
\newcommand{\bpro}{\begin{proof}}
\newcommand{\epro}{\end{proof}}

\makeatletter
\def\vec{\mathop{\operator@font vec}\nolimits}
\makeatother

\begin{document}

\begin{center}
 {\large \bf Orthogonal Series Density Estimation for Complex Surveys}

\bigskip
 Shangyuan Ye, Ye Liang and Ibrahim A. Ahmad \\
 {\small \it Department of Statistics, Oklahoma State University} 

\end{center}
\medskip

\begin{abstract}
We propose an orthogonal series density estimator for complex surveys, where samples are neither independent nor identically distributed. 
The proposed estimator is proved to be design-unbiased and asymptotically design-consistent. The asymptotic normality is proved under 
both design and combined spaces. Two data driven estimators are proposed based on the proposed oracle estimator. 
We show the efficiency of the proposed estimators in simulation studies. A real survey data example is provided for an illustration.
\end{abstract}


\section{Introduction}	
Nonparametric methods are popular for density estimations. Most work in the area of nonparametric density estimation 
was for independent and identically distributed samples. However, both assumptions are violated if the samples are from 
a finite population using a complex sampling design. \cite{Bellhouse99} and \cite{Buskirk99} proposed kernel density estimators (KDE) 
by incorporating sampling weights, and their asymptotic properties were studied by \cite{Buskirk05}. Kernel methods for clustered samples 
and stratified samples were studied in \cite{Breunig01} and \cite{Breunig08}, respectively. 

One disadvantage of the KDE is that all samples are needed to evaluate the estimator. However, in some circumstances, 
there is a practical need to evaluate the estimator without using all samples for confidentiality or storage reasons. 
For example, many surveys are routinely conducted and sampling data are constantly collected. 
Data managers want to publish exact estimators without releasing all original data. In Section 6, we provide a real data example 
from Oklahoma M-SISNet, which is a routinely conducted survey on climate policies and public views. 
The orthogonal series estimators are useful alternatives to KDEs, without needing to release or store all samples. 

The basic idea of the orthogonal series method is that any square integrable function $f$, in our case a density function, 
can be projected onto an orthogonal basis $\{\varphi_j\}$: $f(x) = \sum_{j=0}^{\infty} \theta_j\varphi_j(x)$, where
\be \label{2}
	\theta_j = \int \varphi_j(x)f(x) dx = \mbox{E}(\varphi_j(X)) 
\ee
is called the $j$th Fourier coefficient. Some of the work using orthogonal series was covered in monographs by 
\cite{Efromovich99} and \cite{Tarter93}, among others. \cite{Efromovich10} gave a brief introduction of this method. 
\cite{Walter94} discussed properties of different bases. \cite{Donoho96} and \cite{Efromovich96} studied data driven estimators. 
Asymptotic properties were studied by \cite{Pinsker80} and \cite{Efromovich82}.   

In this paper, we study orthogonal series density estimators (OSDE) for samples from complex surveys. 
To the best of our knowledge, no previous
work has been done on developing OSDE for finite populations. We propose a Horvitz-Thompson type of OSDE, incorporating
sampling weights from the complex survey. We show that the proposed OSDE is design-unbiased and asymptotically 
design-consistent. We further prove the asymptotic normality of the proposed estimator. We compare the lower bound of 
minimax mean integrated squared error (MISE) with the I.I.D. case in \cite{Efromovich82}. We propose two data driven 
estimators and show their efficiency in a simulation study. Finally, we analyze the M-SISNet survey data using the proposed estimation.
All proofs to theorems and corollaries are given in the appendix. 

\section{Notations}
Consider a finite population labeled as $U = \{1, 2, ..., N\}$. A survey variable $x$ is associated with each unit in the finite population. 
A subset $s$ of size $n$ is selected from $U$ according to some fixed-size sampling design $\cfP(\cdot)$. 
The first and second order inclusion probabilities from the sampling design $\cfP(\cdot)$ are $\pi_i = \Pr(i \in s)$ and $\pi_{ij} = \Pr(i,j \in s)$,
respectively. The inverse of the first order inclusion probability defines the sampling weight $d_i = \pi_i^{-1}$, $\forall i \in s$. 

The inference approach used in this paper for complex surveys is the combined design-model-based approach originated in \cite{Hartley75}. 
This approach accounts for two sources of variability. The first one is from the fact that the finite population is a realization from 
a superpopulation, that is, the units $\bfx_U = \{x_1, x_2, ..., x_N\}$ are considered independent random variables with a common distribution 
function $F$, whose density function is $f$. The second one is from the complex sampling procedure which leads to a sample 
$\bfx=\{x_1,x_2, \ldots,x_n\}$. Denote $\bfw=\{w_1, w_2, \ldots,w_n\}$ design variables that determine the sampling weights. 
The sampling design $\cfP(\cdot)$ is embedded within a probability space $(S, \cfJ, P_{\cfP})$. 
The expectation and variance operator with respect to the sampling design are denoted by $\E_{\cfP} (\cdot) = \E_{\cfP} (\cdot \mid \bfx_U)$ 
and $\Var_{\cfP} (\cdot) = \Var_{\cfP} (\cdot \mid \bfx_U)$, respectively. The superpopulation $\xi$, from which the finite population is realized, 
is embedded within a probability space $(\Omega, \cfF, P_\xi)$. The sample $\bfx$ and the design variables $\bfw$ are $\xi$-measurable. 
The expectation and variance operator with respect to the model are denoted by $\E_{\xi}(\cdot)$ and $\Var_{\xi}(\cdot)$, respectively.
Assume that, given the design variables $\bfw$, the product space, which couples the model and the design spaces, 
is $(\Omega \times S, \cfF \times \cfJ, P_{\xi} \times P_{\cfP})$. The combined expectation and variance operators are 
denoted by $\E_C(\cdot)$ and $\Var_C(\cdot)$, where $\E_C(\cdot) = \E_{\xi} [\E_{\cfP} (\cdot \mid \bfx_U)]$ and
$\Var_C(\cdot)= \E_{\xi} [\Var_{\cfP} (\cdot \mid \bfx_U)]+ \Var_{\xi} [\E_{\cfP} (\cdot \mid \bfx_U)]$.

\section{Main Results}
Consider a sample $s = \{x_1, x_2, ...,x_n\}$ drawn from a finite population $\bfx_U$ using some fixed-size sampling design $\cfP(\cdot)$. 
Our goal is to estimate the hypothetical density function $f$ of the superpopulation. Equation (\ref{2}) implies that $\theta_j$ can be
estimated using the Horvitz-Thompson (HT) estimator for the finite population mean
\be \label{coef_est}
	\hat{\theta}_j = N^{-1}\sum_{i=1}^{n} d_i \varphi_j(x_i),
\ee
where $N$ is the finite population size and $d_i = \pi_i^{-1}$ is the sampling weight for unit $i$. 
The HT estimator is a well known design unbiased estimator \citep{horvitz52}. The basis $\{\varphi_j\}$ 
can be Fourier, polynomial, spline, wavelet, or others. Properties of different bases are discussed in \cite{Efromovich10}. 
We consider the cosine basis throughout the paper, which is defined as $\{\varphi_0 = 1, \varphi_j = \sqrt{2}\cos(\pi j x)\}, j = 1,2, \cdots, x \in [0, 1]$. 
Regarding the compact support $[0,1]$ for the density, we adopt the argument in \cite{Wahba81}:`` it might be preferable to assume the 
true density has compact support and to scale the data to interior of $[0,1]$.'' Analogous to \cite{Efromovich99}, 
we propose an orthogonal series estimator in the form 
\be \label{prop_est}
	\hat{f}(x) = \hat{f}(x, \{w_j\}) = 1 + \sum_{j=1}^{\infty} w_j\hat{\theta}_j\varphi_j(x),
\ee
where $\hat{\theta}_j$ is the HT estimator for the Fourier coefficient as in (\ref{coef_est}) and $w_j \in [0,1]$ is a shrinking coefficient. The sequence of $\{w_j\}$ determines the smoothness of the estimator. In Section 4, we consider two choices of $\{w_j\}$ and 
the corresponding data driven estimators, for which only a finite number of $\hat{\theta}_j$ is needed. 
Note that $\theta_0 = \int_{0}^{1} f(x) dx = 1$. If $\bfx_U$ is known for all units in the finite population, we can write the 
population estimator for $f(x)$ as 
\ba
	f_U(x) = f_U(x, \{w_j\}) = 1 + \sum_{j=1}^{\infty}w_j \theta_{U,j}\varphi_j(x),
\ea
where $\theta_{U,j} = N^{-1}\sum_{i=1}^{N}\varphi_j(x_i)$. 

The following theorems and a corollary show properties of our proposed estimator under both design and combined 
spaces. Theorem \ref{thm1} considers unbiasedness and consistency under the design space. 

\begin{theorem} \label{thm1}
Suppose $f \in L_2(\mathbb{R})$, $\delta = N^{-2}\mathop{\sum\sum}_{i \neq k}\frac{\pi_{ik}}{\pi_i \pi_k} - 1 \rightarrow 0$
as $N \rightarrow \infty$,  
and $\sum_{j=1}^{\infty}w_j^2 < \infty$. Then, the estimator $\hat{f}(x, \{w_j\})$ is design-unbiased and asymptotically 
design-consistent for $f_U(x, \{w_j\})$, i.e., 
\ba
	\E_{\cfP} \left[ \hat{f}(x, \{w_j\}) \right] = f_U(x, \{w_j\}) ~\mbox{and}~
	\Gamma_{\cfP} = \Var_{\cfP} \left[ \hat{f}(x, \{w_j\}) \right] \rightarrow 0 ~\mbox{as}~ N \rightarrow \infty.
\ea
\end{theorem}

An intuitive way to understand the condition $\delta\rightarrow 0$ as $N \rightarrow \infty$ is to consider \cite{Hajek64}'s condition:
$\left[\sum_{i \in U}\pi_i (1-\pi_i)\right] (\pi_i\pi_k - \pi_{ik})/\left[\pi_i(1-\pi_i)\pi_k(1-\pi_k)\right] \rightarrow 1$ under that 
$n \rightarrow \infty$ and $N-n \rightarrow \infty$. When $\pi_i = n/N, i\in U$, $\pi_{ik}/(\pi_i \pi_k) \approx 1- n^{-1}(N-n)/N$, 
and hence $\delta\approx -n^{-1}(N-n)/N$. The condition $\delta\rightarrow 0$ is satisfied under that $n \rightarrow \infty$, 
or under \cite{Hajek64}'s condition. Note that the condition is satisfied for $n/N \rightarrow 0$ which is practically plausible. 
We also note that, when the underlying design is a stratified sampling, we restrict our asymptotic framework by assuming 
that the number of strata is finite and fixed.   

The condition $\sum_{j=1}^{\infty}w_j^2 < \infty$ can be easily satisfied by choosing a proper sequence of $\{w_j\}$, which is 
discussed in Section 4. The asymptotic properties here are for the estimator $\hat{f}(x, \{w_j\})$, which we say an `oracle' estimator 
for that $\{w_j\}$ is assumed constant and known. In Section 4, when $\{w_j\}$ is estimated, the estimator becomes
$\hat{f}(x, \{\hat{w}_j\})$, which we say a data driven estimator. 

The following theorem shows the asymptotic normality of the proposed estimator $\hat{f}(x, \{w_j\})$ under the design space.

\begin{theorem} \label{thm2}
Suppose that all assumptions in Theorem \ref{thm1} hold. As $N \rightarrow \infty$,
\ba
	\frac{\hat{f}(x, \{w_j\}) - f_U(x, \{w_j\})}{\hat{\Gamma}_{\cfP}} \xrightarrow{L_{\cfP}} N(0, 1), \label{CLT1}
\ea
where 
\ba
	\hat{\Gamma}_{\cfP} = N^{-1}\sum_{j=1}^{J}w_j^2(1+2^{-1/2}\hat{\theta}_{2j}+\delta \hat{\theta}_j^2)(1+2^{-1/2}\varphi_{2j}(x)).
\ea
\end{theorem}

We then show the asymptotic normality of the proposed estimator $\hat{f}(x, \{w_j\})$ under the combined inference. 
Define a {\it Sobolev Class} of $k$-fold differentiable densities as 
$\cfF(k, Q)$ = $\{f : f(x) = 1 + \sum_{j=1}^{\infty}\theta_j\varphi_j(x), ~\sum_{j=1}^{\infty}(\pi j)^{2k}\theta_j^2 \le Q < \infty\}$, $k \ge 1$. 
Note that for any $f \in \cfF(k, Q)$, $f$ is $1$-periodic, $f^{(k - 1)}$ is absolute differentiable and $f^{(k)} \in L_2(\mathbb{R})$.

\begin{theorem} \label{thm3}
	Suppose that $f \in \cfF(k,Q)$ and all assumptions in Theorem 2 hold. Then, 
	\ba
	\frac{\hat{f}(x, \{w_j\}) - f(x)}{\Var_{C}\left[\hat{f}(x, \{w_j\})\right]} \xrightarrow{L_{C}} N(0, 1) ~\mbox{as}~ N \rightarrow \infty, \label{CLT2}
	\ea
	where $\Var_{C}\left[\hat{f}(x, \{w_j\})\right] = 
	N^{-1} \sum_{j=1}^{J} w_j^2 b_j (1 + 2^{-1/2}\varphi_{2j}(x))$ and $b_j = 2 + 2^{1/2}\theta_{2j} + (\delta - 1)\theta_j^2 + o_N(1)$.
\end{theorem}

The following corollary is a direct result of using Theorem \ref{thm3} and \cite{Efromovich82}. It shows the lower bound 
of the minimax MISE for the proposed estimator $\hat{f}(x, \{w_j\})$ under the Sobolev class. 

\begin{corollary} \label{coro1}
	Let $f \in \cfF(k,Q)$ and $\hat{f}(x,\{w_j\})$ be the estimator in Theorem \ref{thm3}. The lower bound of the minimax MISE, 
	under the combined inference approach, is given by:
	\ba  \label{3}
	R(\cfF) = \inf_{\{w_j\}}\sup_{f \in \cfF(k,Q)} \mbox{MISE}_C\left[ \hat{f}(x,\{w_j\}) \right] \ge N^{-2k/(2k + 1)}P(k, Q, b)(1 + o_N(1)),
	\ea
	where $P(k, Q, b) = Q^{1/(2k+1)}\left\{\frac{k}{\pi (k+1)b}\right\}^{2k/(2k+1)}$ and $b=2$.
\end{corollary}

Remark that this lower bound is of the same form as the I.I.D. case in \cite{Efromovich82}, but with $b=2$ instead of $b=1$. 

\section{Data Driven Estimators}
The choice of shrinking coefficients $\{w_j\}$ is not unique. To get a proper data driven estimator, we start with the oracle estimator (\ref{prop_est}), 
and then obtain $\{w_j\}$ by minimizing the MISE for the oracle estimator. Here, we propose 
two estimators: a truncated estimator and a smoothed truncated estimator, mimicking those in the I.I.D. case.

The truncated estimator, denoted by $\hat{f}_T$, is an estimator with $w_j = 1$ for $j \le J$, and $w_j = 0$ for $j > J$. Alternatively we 
can write $w_j = I_{j \le J}$. Then, only the truncation parameter $J$ needs to be estimated. Notice that the MISE of this estimator is
\ba
	\mbox{MISE}_C \left[\hat{f}(x,\{w_j\}) \right] = \sum_{j=1}^{J} \left[ \Var_C(\hat{\theta}_j) - \theta_j^2 \right] - \int f^2(x) dx.
\ea
Since $\int f^2(x) dx$ is fixed and an unbiased estimator for $\theta_j^2$ is $\hat{\theta}_j^2 - N^{-1}b_j$, a data-driven estimate for $J$ can be 
obtained from
\ba
	\hat{J} = \arg\min \sum_{j=1}^{J} (2N^{-1}\hat{b}_j - \hat{\theta}_j^2),
\ea
where $\hat{b}_j$ is the plug-in estimator of $b_j$. That is, the estimator of the shrinking coefficients can be written as $\hat{w}_j = I_{j \le \hat{J}}$. In practice, the solution is obtained through a numerical search. \cite{Efromovich99} suggests to set the upper bound for $\hat{J}$   
to be $\lfloor 4 + 0.5 \ln n \rfloor$ for the search. Theoretically, the minimum of the MISE can be approximated in the following corollary. 

\begin{corollary} \label{coro2}
	Let $f \in \cfF (k, Q)$, $k > 1/2$. The $\mbox{MISE}$ of $\hat{f}_T$ is minimized when 
	\ba
	J \approx N^{1/(2k+1)} H_1(k,b,c), \label{14}
	\ea	
	and the minimum is approximately
	\ba
	R(\hat{f}_T) = \mbox{MISE}_C(\hat{f}_T(x,\{\hat{w}_j\})) \approx N^{-2k/(2k+1)} H_2(k,b,c), \label{15}
	\ea
	where $H_1(k,b,c) = b^{-1/(2k+1)} \left(\frac{2k+1}{(2k+2)c} \right)^{-1/(2k+1)}$, $H_2(k,b,c) = b^{2k/(2k+1)} \left(\frac{2k+1}{(2k+2)c} \right)^{-1/(2k+1)}$, 
	and $c$ is a constant.
\end{corollary}

One possible modification for $\hat{f}_T$ is to shrink each Fourier coefficient toward zero. We call this estimator the smoothed truncated 
estimator, denoted by $\hat{f}_S$. It is constructed similarly as the truncated estimator, with the first $J$ Fourier coefficients shrunk by multiplying 
the optimal smoothing coefficients $w_j^*$, obtained from the proof of Corollary \ref{coro1}. Mathematically, $\hat{w}_j = \hat{w}_j^* I_{j \le \hat{J}}$, where 
$\hat{w}_j^*=(\hat{\theta}_j^2 - N^{-1}\hat{b}_j)/\hat{\theta}_j^2$ is a direct plug-in estimator for $w_j^*$.  

A potential problem of the nonparametric density estimation is that the estimator may not be a valid density function. A simple modification is to 
define the $L^2$-projection of $\hat{f}_T$ (or $\hat{f}_S$) onto a class of non-negative densities,
$\tilde{f}_T(x) = \max \{0, \hat{f}_T(x) - \mbox{const.}\}$, where the normalizing constant is to make $\tilde{f}_T$ integrate to $1$. 
It has been proved that the constant always exists and is unique \citep{Glad03}.

\section{Simulation}
We compared our proposed estimators with the series estimator that ignores the finite population and sampling designs, through
a Monte Carlo simulation study. We considered estimating density functions for four sampling designs: (1) the simple random sample 
without replacement (SRSWOR), (2) the Poisson sampling, (3) the unequal probability systematic sampling with random start, and (4) the stratified sampling. 
Note that the Poisson sampling and the unequal probability systematic sampling both have a random size and 
hence violates our assumption of fixed size sampling. 
\begin{enumerate}
	\item For the SRSWOR, we considered two superpopulations: the standard normal distribution $\mbox{N}(0, 1)$ 
		and a mixture normal distribution $0.4\mbox{N}(-1, 0.5)+0.6\mbox{N}(1, 1)$. 
	\item For the Poisson sampling, we considered the same two superpopulations as in (1). 
		We specified the expected sample size for the Poisson sampling to be $n$, 
		with inclusion probabilities $\pi_i \propto \log\{\max(x_i+5,1)\}$.
	\item For the unequal probability systematic sampling, we considered the same two superpopulations as in (1). 
		We specified the inclusion probabilities as $\pi_i \propto \log\{\max(x_i+5,1)\}$. 
	\item For the stratified sampling, we considered two superpopulations: a two-component mixture normal
		$0.4\mbox{N}(-1, 0.5)+0.6\mbox{N}(1, 1)$ and a three-component mixture normal
		$0.3\mbox{N}(-1, 0.15)+0.4\mbox{N}(0, 0.15)+0.3\mbox{N}(1, 0.15)$. We designed two strata for 
		the two-component mixture and three strata for the three-component mixture. A proportional stratified sampling is used.
\end{enumerate}
For all cases, we considered a finite population of size $N=1,000$ drawn from each of the superpopulations. 
We repeated drawing the finite population for $m_1=100$ times. For each of the finite population, we drew samples according to the sampling design, 
with increasing sample sizes: $n = 20, 40, 60, 80$ and $100$. The replication number for each finite population is $m_2=10,000$. 
The performance of estimators is measured by a Monte Carlo approximation of the MISE: 
\ba
	\mbox{MISE}_{\mbox{MC}} (\tilde{f}) = \int \frac{1}{m_1m_2}\sum_{i=1}^{m_1}\sum_{j=1}^{m_2} \left[\tilde{f}_{ij}(x) - f(x) \right]^2 dx.
\ea

The results of the simulation study are shown in Table 1. In general, the I.I.D. series estimator, which ignores the sampling design, performs the worst in nearly all cases. It confirms the necessity of incorporating sampling weights into the series estimator for a complex survey. 
Moreover, the improvement of the proposed estimators is even bigger in the mixture case than the standard normal case. 
Lastly, the difference between the smoothed truncated estimator and the truncated estimator is quite small. 

\section{Oklahoma M-SISNet Survey}
\begin{figure}
	\begin{center}
		\includegraphics[scale=0.8, angle=270]{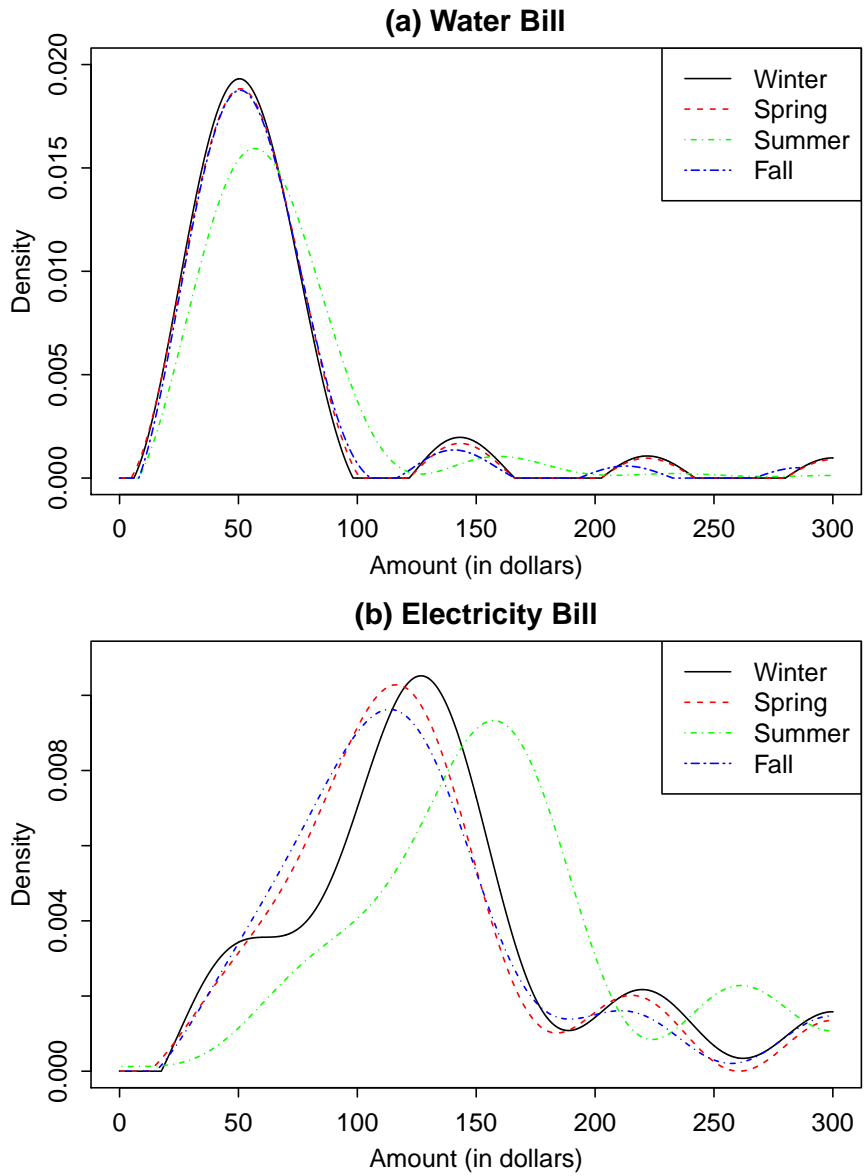}
		\caption{OSDEs of the electricity bill and the water bill for seasonal waves in 2015.}
		\label{Figure1}
	\end{center}
\end{figure}

The Oklahoma Weather, Society and Government Survey conducted by Meso-Scale Integrated Sociogeographic Network (M-SISNet) 
measures Oklahomans' perceptions of weather in the state, their views on government policies and societal issues
and their use of water and energy. The survey is routinely conduced at the end of each season. Until the end of 2016, 
12 waves of survey data have been collected. It is desired that estimates can be obtained without constantly pulling out 
the original data. The sampling design has two separated phases. In Phase I, a simple random sample of size $n = 1,500$ 
is selected from statewide households. In Phase II, a stratified oversample is selected from five special study areas: 
Payne County, Oklahoma City County, Kiamichi County, Washita County and Canadian County. In each stratum,  
the sample size is fixed to be $200$. The second phase can be viewed as a stratified sampling over the entire state with
six strata: $n_1 = \cdots = n_5 = 200$ and $n_6=0$, where the sixth stratum contains households not in the five special study areas. 
This design with oversampling is not a typical fixed-size complex survey. The first-order inclusion probabilities 
are approximately $\pi_{hi} = n_h/N_h + n/N$, for $i=1,\ldots,N_h$ and $h=1,\ldots,6$. Note that for units not in the five areas, 
this inclusion probability is simply $n/N$. We presents OSDEs for two continuous variables for illustration: the
monthly electricity bill and the monthly water bill. Figure \ref{Figure1} shows OSDEs of the two variables for all seasons in 2015. 
The upper panel of Figure \ref{Figure1} shows the water bill distribution and the lower panel shows the electricity bill distribution. 
We clearly see the difference between four seasons for the consumption of water and electricity. For example, the summer electricity 
bill is much higher than other seasons. Notice that those densities appear to be multimodal, while our simulation studies suggested 
that our proposed estimators enjoy a remarkable improvement in cases with mixture densities. 

\section{Conclusion}
In this paper, we propose a Horvitz-Thompson type orthogonal series estimator for density estimation under complex survey designs.
The estimator is shown to possess certain asymptotic properties and outperforms the ordinary orthogonal series estimator without sampling weights.  
It is of practical use when an exact nonparametric estimator needs to be released without releasing the original data. 
In the future, it is worth investigating the orthogonal series estimation under survey designs when auxiliary variables are present 
or nonresponse occurs. 

\section*{Acknowledgement}
This research is partially supported by National Science Foundation under Grant No. OIA-1301789. 
We thank an associate editor and a referee for their constructive comments. 

\section*{Appendix}

\subsection*{Proof of Theorem 1}
\begin{proof}
	We first show that $\hat{f}(x, \{w_j\})$ is design-unbiased:
	\ba
	\E_{\cfP} \left[ \hat{f}(x, \{w_j\}) \right] &=& \E_{\cfP} \left[1 + \sum_{j=1}^{\infty} w_j\hat{\theta}_j\varphi_j(x) \right] \\
	&=&
	1 + \sum_{j=1}^{\infty} w_j\E_{\cfP}(\hat{\theta}_j)\varphi_j(x) \\ 
	&=& 
	1 + \sum_{j=1}^{\infty}w_j \theta_{U,j}\varphi_j(x) \\ &=& f_U(x, \{w_j\}).
	\ea
	It remains to show that $\hat{f}(x, \{w_j\})$ is asymptotically design-consistent, that is, the design-variance of 
	$\hat{f}(x, \{w_j\})$ approaches zero in the limit. We need the simple fact that 
	\ba
	\varphi_j^2(x) = [\sqrt{2}\cos(\pi j x)]^2 = 1 + \cos(\pi 2j x) = 1 + 2^{-1/2}\varphi_{2j}(x).
	\ea
	Then, we have
	\ba
	\Gamma_{\cfP} 
	&=& \Var_{\cfP} \left[ 1 + \sum_{j=1}^{\infty}w_j\hat{\theta}_j\varphi_j(x) \right] \\ 
	&=& \sum_{j=1}^{\infty}w_j^2\varphi_j^2(x)\Var_{\cfP} (\hat{\theta}_j) \\ 
	&=& \sum_{j=1}^{\infty} w_j^2 \left[1 + 2^{-1/2}\varphi_{2j}(x) \right] N^{-2} \Var_{\cfP}\left[\sum_{i=1}^{n}d_i\varphi_j(x_i) \right],
	\ea
	and
	\ba
	\Var_{\cfP} \left[\sum_{i=1}^{n}d_i\varphi_j(x_i)\right] 
	&=& \Var_{\cfP} \left[\sum_{i=1}^{N}I_id_i\varphi_j(x_i)\right] \\ 
	&=& \E_{\cfP}\left[\sum_{i=1}^{N}I_id_i\varphi_j(x_i)\right]^2 - \left\{\E_{\cfP}\left[\sum_{i=1}^{N}I_id_i\varphi_j(x_i)\right]\right\}^2 \\ 
	&=& \sum_{i=1}^{N}\E_{\cfP}(I_i^2)d_i^2\E_{\cfP} \left[\varphi_j^2(x_i)\right] 
	+ \mathop{\sum\sum}_{i \neq k} \pi_{ik} d_i d_k \E_{\cfP} \left[\varphi_j(x_i)\right] \E_{\cfP}\left[\varphi_k(x_k)\right] \\
	& & - \left\{\sum_{i=1}^{N} \E_{\cfP}(I_i) d_i \E_{\cfP}\left[\varphi_j(x_i)\right] \right\}^2 \\
	&=& \sum_{i=1}^{N} \E_{\cfP} \left[1 + 2^{-1/2}\varphi_{2j}(x_i)\right] + \mathop{\sum\sum}_{i \neq k}\frac{\pi_{ik}}{\pi_i \pi_k} \theta_{U,j}^2 - N^2\theta_{U,j}^2 \\ 
	&=& N (1 + 2^{-1/2}\theta_{U,2j}) + N^2 \delta \theta_{U,j}^2 \\ 
	&\le& NM_1 + N^2 \delta M_2, 
	\ea
	where $1 + 2^{-1/2}\theta_{U,2j} \leq M_1 < \infty$ and $\theta_{U,j}^2 \leq M_2 < \infty$ for every $j$.
	
	Hence, $\Gamma_{\cfP} \le 2(N^{-1} M_1 + \delta M_2) \sum_{j=1}^{\infty} w_j^2 \rightarrow 0$ as $N \rightarrow \infty$.
\end{proof}

\subsection*{Proof of Theorem 2}
\begin{proof}
	By the definition of $\hat{\theta}_j$ and $\theta_{U,j}$, we have 
	\ba
	\hat{f}(x, \{w_j\}) &=& 1 + \sum_{j=1}^{\infty} w_j \hat{\theta}_j \varphi_j(x) \\ &=& 1 + \sum_{i=1}^{N} I_i d_i \sum_{j=1}^{\infty} w_j \varphi_j(x) \varphi_j(x_i),
	\ea
	and 
	\ba 
	\E \left[ I_i d_i \sum_{j=1}^{\infty} w_j \varphi_j(x) \varphi_j(x_i) \right] &=& \sum_{j=1}^{\infty} w_j \varphi_j(x) \E\left[ \varphi_j(x_i)\right] \\ &=& \sum_{j=1}^{\infty} w_j \theta_{U,j} \varphi_j(x).
	\ea
	Also, from the proof of Theorem 1, we have
	\ba
	\Var \left[ I_i d_i \sum_{j=1}^{\infty} w_j \varphi_j(x) \varphi_j(x_i) \right] &=& \sum_{j=1}^{\infty} w_j^2 (1 + 2^{-1/2}\theta_{U,2j} + \delta \theta_{U,j}^2) \\ &\le& B\sum_{j=1}^{\infty} w_j^2 < \infty ~ \text{by assumption.}
	\ea
	Therefore, by the Lindeberg-L$\acute{e}$vy central limit theorem, we have
	\be
	\frac{\hat{f}(x, \{w_j\}) - f_U(x, \{w_j\})}{\Gamma_{\cfP}} \xrightarrow{L_{\cfP}} N(0, 1). \label{C1}
	\ee
	It remains to show that $\hat{\Gamma}_{\cfP}$ is consistent for $\Gamma_{\cfP}$ under design, or equivalently,
	\be
	|\hat{\Gamma}_{\cfP} - \Gamma_{\cfP}| \xrightarrow{P_{\cfP}} 0, ~\text{as}~ n \rightarrow N. \label{C2}
	\ee
	Condition (\ref{C2}) can be proved by using the facts that $\hat{\theta}_j$ is design unbiased and
	$\E(\hat{\theta}_j^2) = \theta_j^2 + \Var(\hat{\theta}_j) \rightarrow \theta_j^2$ as $n \rightarrow N$.
	
	Then, Theorem 2 is proved by using the equations (\ref{C1}) and (\ref{C2}) in conjunction with Slutsky's theorem.
\end{proof}

\subsection*{Proof of Theorem 3}
\begin{proof}
	Since $f_U(x, \{w_j\})$ is the standard OSDE from an I.I.D. sample which is the finite population, then
	\be
	\frac{f_U(x, \{w_j\}) - f(x)}{\Var_{\xi}\left[f_U(x, \{w_j\})\right]} \xrightarrow{L_{\xi}} N(0, 1). \label{CLT3}
	\ee
	The asymptotic distribution of the I.I.D. OSDE under Sobolev class is obtained from \cite{Efromovich99}, Chapter 7. 
	Also,
	\be
	\Var_C\left[\hat{f}(x, \{w_j\})\right] &=& \sum_{j=1}^{J} \Var_C \left[w_j\hat{\theta}_j\varphi_j(x)\right] \nonumber \\ &=& \sum_{j=1}^{J} w_j^2 (1 + 2^{-1/2}\varphi_{2j}(x)) \Var_C(\hat{\theta_j}) \label{var}
	\ee 
	Next, we calculate the variance of $\hat{\theta}_j$ by using Theorem 1:
	\be
	\Var_C(\hat{\theta}_j) &=& \E_{\xi} \left[\Var_{\cfP}(\hat{\theta}_j) \right] + \Var_{\xi} \left[\E_{\cfP}(\hat{\theta}_j)\right] \nonumber \\ 
	&=& \E_{\xi} \left[N^{-1}(1 + 2^{-1/2}\theta_{U,2j} + \delta\theta_{U,j}^2) \right] + \Var_{\xi}(\theta_{U,j}) \nonumber \\ 
	&=& N^{-1} \left[1 + 2^{-1/2}\theta_{2j} + \delta\E_\xi(\theta_{U,j}^2) \right] + \Var_\xi(\theta_{U,j}) \label{7}
	\ee
	Then, we evaluate $\E_{\xi}(\theta_{U,j}^2)$ and $\Var_{\xi}(\theta_{U,j})$ separately. Based on a standard result in the 
	I.I.D. case, we have
	\be
	\Var_{\xi}(\theta_{U,j}) = N^{-1}(1 + 2^{-1/2}\theta_{2j} - \theta_j^2) \label{8}
	\ee
	and
	\be
	\E_{\xi}(\theta_{U,j}^2) &=& \E_{\xi}^2(\theta_{U,j}) + \Var_{\xi}(\theta_{U,j}) \nonumber \\ 
	&=& N^{-1}(1 + 2^{-1/2}\theta_{2j} - \theta_j^2) + \theta_j^2. \label{9}
	\ee
	Then, plug equations (\ref{8}) and (\ref{9}) into (\ref{7}), we have
	\be
	\Var_C(\hat{\theta}_j) = N^{-1} \left[2 + 2^{1/2}\theta_{2j} + (\delta - 1)\theta_j^2 + o_N(1) \right] = N^{-1}b_j. \label{10}
	\ee
	Hence, plug (\ref{10}) into (\ref{var}) we can get the variance of $\hat{f}$ under the combined inference approach. \\
	Finally, apply Theorem 5.1 in \cite{Bleuer05}, Theorem 3 is proved. 
\end{proof}

\subsection*{Proof of Corollary 1}
\begin{proof}
	The proof is similar to \cite{Efromovich82}. We sketch the steps as follows.
	We first evaluate the linear minimax MISE for the functions in the Sobolev class defined above. That is, we optimize 
	$w_j^*$'s that minimize $\mbox{MISE}_C(\hat{f})$. 
	Notice that $\E_C(\hat{\theta}_j) = \E_{\xi} [\E_{\cfP}(\hat{\theta}_j)] = \E_{\xi}(\theta_{U,j}) = \theta_j$ 
	implying that $\hat{\theta}_j$ is an unbiased estimator of $\theta_j$. Therefore,
	\be 
	\mbox{MISE}_C \left[\hat{f}(x, \{w_j\}) \right] 
	&=& \E_C \left[ \int (f - \hat{f})^2 \right] \nonumber \\ 
	&=& \sum_{j=1}^{\infty} \left\{w_j^2 \left[\Var_C(\hat{\theta}_j) + \theta_j^2\right] - 2w_j\theta_j^2 + \theta_j^2 \right\}. \label{4}
	\ee
	A straightforward calculation yields that
	\be \label{5}
	w_j^* = \frac{\theta_j^2}{\theta_j^2 + \Var_C(\hat{\theta}_j)}. 
	\ee
	Plug equation (\ref{5}) into (\ref{4}), 
	\be
	R_L(\cfF) &=& \inf_{\{w_j\}}\sup_{f \in \cfF(k,Q)} \mbox{MISE}_C \left[\hat{f}(x,\{w_j\}) \right] \nonumber \\
	&\ge& \sup_{f \in  \cfF(k,Q)} \sum_{j=1}^{\infty} \frac{\theta_j^2 \Var_C(\hat{\theta}_j)}{\theta_j^2 + \Var_C(\hat{\theta}_j)}, \label{6} 
	\ee 
	where $\Var_C(\hat{\theta}_j)$ is of the form (\ref{10}).
	Plug (\ref{10}) into (\ref{6}), 
	and use the Lagrange multiplier to show that the maximum of (6) is attained at
	\be \label{11}
	\theta_j^2 = N^{-1} (\mu / (\pi j)^k - b_j)_+, 
	\ee
	where $\mu$ is determined by the constraint $\sum_{j=1}^{\infty}(\pi j)^{2k}\theta_j^2 \le Q$. Plug equation (\ref{11}) back 
	to (\ref{6}), we obtain
	\ba
	R_L(\cfF) \ge N^{-2k/(2k + 1)}P(k, Q, b).
	\ea
	\cite{Pinsker80} shows that for Sobolev ball $\cfF$, the linear minimax risk is asymptotically equal to the minimax risk, that is,
	$R(\cfF) = R_L(\cfF)(1 + o_N(1))$. Therefore Corollary 1 is proved.
\end{proof}

\subsection*{Proof of Corollary 2}
\begin{proof}
	Let $\hat{w}_j = I_{j \le J}$. Plug equation (\ref{10}) into (\ref{4}), we have
	\be
	R(\hat{f}_T) = N^{-1}\sum_{j=1}^{J} b_j + \sum_{j = J + 1}^{\infty} \theta_j^2 \approx N^{-1}bJ + \sum_{j = J + 1}^{\infty} \theta_j^2. \label{12}
	\ee
	Notice that for $f \in \cfF(k,Q)$. By a straightforward calculation, we have $\theta_j^2 = cj^{-2(k+1)}$ \citep{Efromovich99}. Therefore,
	\be
	\sum_{j = J + 1}^{\infty} \theta_j^2 \approx c \int_{J}^{\infty} j^{-2(k+1)} dj = \frac{c}{2k+1} J^{-2k - 1}. \label{13}
	\ee
	Plug (\ref{13}) into (\ref{12}) and optimize $J$, Corollary 2 is proved.
\end{proof}

\renewcommand{\baselinestretch}{1.1}


\newpage
\begin{table}[ht]
\caption{Monte Carlo approximation of MISE for four sampling designs and two superpopulations. The finite population
	size is $N=1,000$. The replication size of the finite population is $m_1 = 100$, and the replication size of the sample is $m_2 = 10,000$. Three estimators are compared: the truncated estimator,
	the smoothed estimator and the series estimator ignoring finite population and sampling design (I.I.D.).}
\begin{center}
\begin{tabular}{ c|c|c|c|c|c|c }
	\hline
	\hline
	\multicolumn{7}{c}{SRSWOR} \\
	\hline
	& \multicolumn{3}{|c|}{ Standard Normal } & \multicolumn{3}{c}{ Mixture Normal } \\
	\hline
	n & Truncated & Smoothed & I.I.D. & Truncated & Smoothed & I.I.D. \\
	\hline
	20 & 0.0232 & \bf 0.0220 & 0.0290 & 0.0498 & \bf 0.0480 & 0.0535\\
	40 & 0.0150 & \bf 0.0140 & 0.0157 & \bf 0.0311 & 0.0318 & 0.0388\\
	60 & 0.0116 & \bf 0.0109 & 0.0121 & \bf 0.0226 & 0.0234 & 0.0335\\
	80 & 0.0094 & \bf 0.0089 & 0.0100 & \bf 0.0173 & 0.0180 & 0.0219\\
	100 & \bf 0.0064 & 0.0067 & 0.0071 & \bf 0.0134 & 0.0139 & 0.0139\\
	\hline
	\hline
	\multicolumn{7}{c}{Poisson Sampling} \\
	\hline
	& \multicolumn{3}{|c|}{ Standard Normal } & \multicolumn{3}{c}{ Mixture Normal } \\
	\hline
	n &Truncated & Smoothed & I.I.D. & Truncated & Smoothed & I.I.D. \\
	\hline
	20 & 0.0328 & 0.0317 & \bf 0.0210 & 0.0346 & \bf 0.0343 & 0.0437\\
	40 & 0.0158 & \bf 0.0153 & 0.0157 & \bf 0.0193 & 0.0194 & 0.0341\\
	60 & 0.0123 & \bf 0.0117 & 0.0136 & \bf 0.0158 & 0.0159 & 0.0295\\
	80 & 0.0096 & \bf 0.0091 & 0.0120 & \bf 0.0133 & 0.0135 & 0.0262\\
	100 & 0.0076 & \bf 0.0071 & 0.0108 & \bf 0.0115 & 0.0117 & 0.0234\\
	\hline
	\hline
	\multicolumn{7}{c}{Unequal Probability Systematic Sampling} \\
	\hline
	& \multicolumn{3}{|c|}{ Standard Normal } & \multicolumn{3}{c}{ Mixture Normal } \\
	\hline
	n &Truncated & Smoothed & I.I.D. & Truncated & Smoothed & I.I.D. \\
	\hline
	20 & 0.0277 & 0.0268 & \bf 0.0218 & 0.0311 & \bf 0.0309 & 0.0419\\
	40 & 0.0152 & \bf 0.0147 & 0.0165 & \bf 0.0183 & 0.0185 & 0.0326\\
	60 & 0.0125 & \bf 0.0118 & 0.0139 & \bf 0.0140 & 0.0143 & 0.0290\\
	80 & 0.0089 & \bf 0.0083 & 0.0128 & \bf 0.0129 & 0.0132 & 0.0275\\
	100 & 0.0058 & \bf 0.0054 & 0.0108 & \bf 0.0125 & 0.0127 & 0.0226\\
	\hline
	\hline
	\multicolumn{7}{c}{Stratified Sampling} \\
	\hline
	& \multicolumn{3}{|c|}{ Two Strata } & \multicolumn{3}{c}{ Three Strata } \\
	\hline
	n &Truncated & Smoothed & I.I.D. & Truncated & Smoothed & I.I.D. \\
	\hline
	20 & 0.0415 & \bf 0.0409 & 0.0739 & 0.2847 & \bf 0.2826 & 0.3106\\
	40 & 0.0231 & \bf 0.0230 & 0.0688 & 0.2731 & \bf 0.2718 & 0.3309\\
	60 & 0.0181 & \bf 0.0180 & 0.0672 & 0.0426 & \bf 0.0419 & 0.1132\\
	80 & \bf 0.0142 & \bf 0.0142 & 0.0675 & 0.0412 & \bf 0.0406 & 0.1175\\
	100 & \bf 0.0128 & 0.0129 & 0.0601 & \bf 0.0381 & 0.0395 & 0.1128\\
	\hline	
\end{tabular}
\end{center}
\label{figure 1}
\end{table}%

\newpage

\end{document}